Linking stem density, stem size and spatial arrangement: an approach to model discriminative self-thinning in even-aged forest stands

Vladimir L. Gavrikov[1,*], Rem G. Khlebopros[1,2]

[1]Siberian Federal University, Institute for Economics, Management and Environmental Studies Chair of Ecology and Environmental Studies, Krasnoyarsk, Russian Federation

[2]International scientific research center for extreme conditions of organism KSC SB RAS, Krasnoyarsk, Russian Federation

*Corresponding author:

79 pr. Svobodnyi, Krasnoyarsk, 660041, Russian Federation

Phone: (+7 913 042 4304), e-mail: vgavrikov@sfu-kras.ru

**Abstract**

To model discriminative, i.e. competition induced, self-thinning in even-aged forest stands a concept has been explored that discriminative mortality alters spatial arrangement of trees which in turn alters the mortality. Function of density was suggested to be a balance of initial density minus mortality that was dependent on initial spatial arrangement and mean horizontal size in the forest stand. Assuming initial spatial arrangement to be aggregation and performing normalizations gave the self-thinning function depending on only two parameters, initial stem density and maximal horizontal size (represented by stem diameter). Normalizations included integral of probability of trees to collide to be unity and stem density at maximal size to be zero. The self-thinning function obtained has been shown to successfully capture non-linear self-thinning dynamics in Douglas-fir long-term experiments in which competition-induced mortality prevailed.

Key words: modeling competition, discriminative mortality, self-thinning, even-aged stands, spatial arrangement, Douglas-fir.

**Introduction**

Studies of population dynamics in trees have been rather numerous and topical for decades. Since some early works (Mohler et al., 1978; Buzykin, Khlebopros, 1981) it has been suggested to differentiate between random mortality and discriminative, competition-induced, mortality in the course of self-thinning. The former takes mostly place in the very early stages of stand growth and after growth has stopped. The latter prevails all the time when there is intensive competition among fast growing trees in dense stands.

Data on how stem density evolves under various conditions are abundant and usually reflected in yield tables. Still, receiving of a law that explains self-thinning even in simple even-aged forest stands has been quite a challenge for researchers. Since long ago, forestry practitioners (see, e.g. Frothingham, 1914) have known that mean diameter in a forest stand was inversely related to stem density in the stand. This observation is true for a given age of different stands, a static approach, but also for a particular stand that grows and ages with time, a dynamic approach. Both approaches use the term 'self-thinning' but they differ in treating data.

The static approach is mostly associated with such famous concepts as Reineke's rule (Reineke, 1933) and '–3/2' self-thinning rule (Yoda et al., 1963). The underlying idea of the concepts is that average sizes of plants in dense populations, on the one hand, and the numbers of plants in the populations, on the other, are linked to each other by a power function with a negative exponent. Then, plotted in log–log coordinates the function is a straight line, with the exponent being the slope of it. Originally, the authors of the rules suggested that the slopes of the self-thinning lines to be constant and universal among species and stands, –1,602 and –3/2 for Reineke's rule and Yoda's rule, correspondingly. The suggestions have been criticized both empirically and theoretically (von Gadow, 1986; Zeide, 1987). Subsequent studies have shown that the exponents had rather species-specific values. Recent analyses on the matter may be found in a number of publications (Pretzsch, Biber, 2005; Pretzsch, 2006; Vanclay, Sands, 2009;

Larjavaara, 2010; Gavrikov, 2015). Nevertheless, the self-thinning rules were found to be important empirical generalizations and also evoked fruitful researches on potential (maximal) density of forest stands (Sterba, 1987; Sterba, Monserud, 1993, 1995; Vospernik, Sterba, 2015).

The dynamic approach, not contradicting to the static one, puts a focus on the process in which an individual forest stand looses trees due to mortality as its trees grow larger and compete with each other. Self-thinning as a dynamic process is a frequent topic of research. Sometimes it involves quite complicated mathematical descriptions. For example, Pittman and Turnblom (2003) used a system of three differential equations including 15 coefficients to show that each forest stand follows its own self-thinning curve depending on allometric relations between various parameters, with the relation being individual to the stand.

Another branch of the approach seeks to derive self-thinning laws from interrelationships at individual tree level. A key point here is to understand the mechanisms of individual trees competition that may explain the time course of density falling. One of early attempts to get the dynamic law of self-thinning was a work by Slatkin and Anderson (1984). The authors developed a two-dimensional model of competition for space in which growing circles collided as they grew and in this sense 'competed' for space. When two circles collided one of them disappeared, this simulated mortality in their 'population'. This kind of instant mortality was an obvious simplification which allowed a mathematical analytical study while the model still reflected growth and competition of organisms with fixed positions. As a result, Slatkin and Anderson received a function describing self-thinning in the form

$$F(R) = \frac{1}{1 + 2\pi R^2 D},$$

where $F(R)$ is a function that gives probability for a random circle to reach radius $R$ and $D$ is initial density of the circles. Probability is a less observable matter than density, so Gavrikov (1995) tried to use Slatkin and Anderson's growing circle population to derive a function of self-thinning. The function linked population density $D(n)$ at a discreet time moment $n$, initial density of circles $D_0$, discreet growth rate $\Delta R$ and a variable $G$ describing the probability of circles to collide as their grew by $\Delta R$:

$$D(n) = D_0(1 - G \cdot \Delta R)^n. \qquad (1)$$

It is noteworthy that the approximate solution of the model required an explicit definition of how the circles distributed over the two-dimensional space. For the particular form of the function $D(n)$ (1) it was necessary to suppose that the circles centers were randomly distributed.

Information on spatial arrangement should be quite of importance for modeling of population of such organisms like trees. These organisms cannot avoid collision with competitors at short range and have to either win in competition or die off. It is spatial arrangement that will determine how frequent collisions with neighboring individuals take place.

Trees are not always randomly distributed in natural population. It has been shown, for example, that trees in Scots pine (Pinus sylvestris L.) even-aged stands show strong clumping at younger ages and with age they approach a random distribution for centers of stems and a uniform distribution for centers of crowns (Gavrikov et al., 1993). In uneven-aged populations of Siberian fir (Abies sibirica Ledeb.), younger trees are in clumps and completely grown trees are arranged close to random pattern (Gavrikov, Stoyan, 1995). Natural disturbances, such as catastrophic winds, often have the effect that new generation of trees appears in clumps (Szmyt, Dobrowolska, 2016).

In recent years, interest has been shown in using of spatial moments techniques for modeling of various biological processes in general (Simpson, Baker, 2015) and forest stand structure in particular (Adams et al., 2013). Spatial moment's theory is quite an in-depth framework that includes complex mathematical descriptions. Adams et al. (2013) developed a model describing first and second spatial moment dynamics in a forest stand, with neighbor-dependent plant growth being explicitly incorporated in the model.

Aims of the study were to i) develop a mathematically tractable approach that allows one to incorporate spatial arrangement into a function describing self-thinning of an even-aged stand and ii) test the function against real field data of self-thinning in even-aged forest stands.

**Data**

To test the derived formulas against field data a number of openly published datasets were used. All of them belong to the series of experiments that were run by US Forest Service since 1960s. The experiments are known under the name of Levels-of-growing-stock Cooperative Study in Douglas-fir (Pseudotsuga menziesii (Mirb.) Franco). The experiments were thought to study reaction of forest stands to various rates and schemes of artificial thinning. Apart from plots subjected to thinnings, the experiments included a number of control plots in which no thinning interventions took place and which were thought to serve as a comparison for experimental plots. These control plots are therefore convenient objects to observe a natural-like self-thinning process. All the plots were measured at approximately regular time intervals and the data were published as working reports.

For this study four datasets were available: Hoskins (Marshall, Curtis, 2001), Skykomish Study and Clemons Study (King et al. 2002), Iron Creek Study (Curtis, Marshall, 2009).

The fitting of the data against the model developed below was performed with the help of Statistica 6 software.

**Results and discussion**

*Model development*

Suppose, spatially fixed organisms like trees are distributed over large enough homogenous space. Initial number of trees may be given by a variable $D_0$. All the trees have horizontal dimensions, most popular of them being stem diameter and crown diameter. For the purposes of modeling, the horizontal measure of trees is represented by a mean stem diameter $r$ for the whole population; it is supposed however that stem diameter and crown diameter are well correlated with each other so that only one of them can be used. Dimensions of trees are limited by the maximal size that trees can achieve that is given in the model as a value $R_{max}$. In the course of growth, i.e. as $r$ increases, trees may 'collide', i.e. to come to touch, which supposes competition for space. Of course, trees do not 'collide' by stems but rather by crowns. The probability to collide, mean dimension being $r$, may be described by a function $G(r)$ that will be specified below. Some individuals that have collided are to die off due to competition-induced mortality. This kind of mortality may be described by a variable $\gamma$.

Thus, a balance equation of self-thinning in such a forest stand may be given by

$$D(r) = D_0 - \gamma \cdot D_0 \int_0^r G(\rho) d\rho \,. \qquad (2)$$

It follows from (2) that variable $\gamma$ scales the number of collided individuals and should have the dimension $r^{-1}$. The function $G(r)$ plays a key role in the description of competition-induced mortality. A definite form of $G(r)$ will be below suggested and the equation (2) will be analyzed with its help.

For the model being developed, it is crucial to define a measure that would help to predict probability of collisions as trees grow in horizontal size. In spatial statistics, one of popular measures, pair correlation function (also known as Ripley's function (Ripley, 1977)), might be very useful for this purpose. In its basic formulation, pair correlation function $g(x)$ is the density of pairs of objects distributed over space and separated by distance $x$. The function $g(x)$ may be normalized in such a way that for Poisson process it will return unity for every $x$, i.e. $g(x) = const = 1$. It means that every distance $x$ is equally probable and the graph $g(x)$ is a horizontal line. Frequent deviations from Poisson distribution are aggregation and repulsion. For objects arranged in aggregations, $g(x)$ will be greater than unity for small $x$ but will tend to unity for larger $x$. For repulsing objects, $g(x)$ will be smaller than unity for small $x$ but, again, will tend to unity for larger *x*. Depending on normalization procedure, other constants than unity may serve as a reference point.

Unfortunately, the pair correlation function in its spatial statistics definition is rather difficult to use for analytical modeling pursued here. This is because the pair correlation function is based on explicit definition of all the individual objects in the population. Such a microscopic view is completely just but makes the analysis cumbersome. A way out is to suggest a macroscopic approximation for pair correlation function that would behave 'like' the function but have small number of tractable parameters.

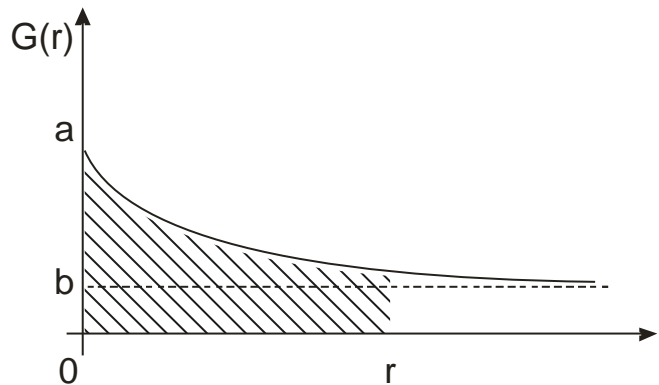

Figure 1. Graphical representation of function *G(r)* in the form $G(r) = \dfrac{a+br}{1+r}$ (see (3)) for the case of aggregated spatial arrangement, *a* and *b* being parameters of the function. The hatched area is integral of *G(r)* from *0* to *r*.

Function *G(r)* is suggested therefore to consider in the form

$$G(r) = \frac{a+br}{1+r}, \qquad (3)$$

where *a* and *b* are parameters and *r* is the measure of horizontal size. Function (3) simulates probability to collide for all the basic types of spatial arrangements. Figure 1 shows the form of function (3) for the case *a* > *b*, which corresponds to aggregated spatial arrangement (maximal values at small *r*). For the case of arrangement of repulsing objects the inequality *a* < *b* should take place (minimal values at small *r*). The case $a = b$ means that $G(r) = const = b$, which reflects that there is equal probability to collide as in Poisson arrangement.

Integration of (3) from *0* to *r* according to (2) gives

$$\int_0^r \frac{a+b\rho}{1+\rho} d\rho = (a-b)\ln(1+r) + br \qquad (4)$$

and therefore the basic equation (2) with substitution (4) can be rewritten as

$$D(r) = D_0 - \gamma \cdot D_0 [(a-b)\ln(1+r) + br]. \qquad (5)$$

Finally, a few normalizing procedures are necessary in order to use *G(r)* as probability in computations.

First, the integral (4) taken from *0* to $R_{max}$ should be less or equal to unity (see fig. 1). It is suggested here to use the normalization in the form

$$(a-b)\ln(1+R_{max}) + bR_{max} = 1, \qquad (6)$$

which means that by the moment of achieving size $R_{max}$ trees will surely collide. It is clear that such a supposition is valid for dense enough populations subjected to competition-induced mortality.

Then the condition (6) gives an opportunity to express *a* though *b* as

$$a = \frac{1 - b(R_{max} - \ln(1+R_{max}))}{\ln(1+R_{max})}. \qquad (7)$$

Second, because the model consideration is limited by $R_{max}$ it would be natural to limit function *G(r)* by $R_{max}$ as well. The condition looks like $G(R_{max}) = 0$, which means that after the moment of forest stand achieves size of $R_{max}$ no more collisions are possible. According to (3), the condition is

$$a = -b \cdot R_{max}. \qquad (8)$$

Obviously, relations (7) and (8) can be solved as a system of equations so that variables *a* and *b* can be then expressed through $R_{max}$. Therefore, in the basic equation (5), the normalizations (7) and (8) allow one to substitute two less observable or unknown beforehand variables (*a* and *b*) through a more observable variable ($R_{max}$).

Through solving of (7) and (8) one gets

$$b = \frac{1}{R_{max}(1 - \ln(1 + R_{max})) - \ln(1 + R_{max})}$$

$$a = \frac{R_{max}}{\ln(1 + R_{max}) - R_{max}(1 - \ln(1 + R_{max}))} \qquad (9)$$

and therefore the final form of (5) will look as

$$D(r) = D_0 - \gamma \cdot D_0 \left[ \left( \frac{1 + R_{max}}{\ln(1 + R_{max}) - R_{max}(1 - \ln(1 + R_{max}))} \right) \ln(1 + r) + \frac{1}{R_{max}(1 - \ln(1 + R_{max})) - \ln(1 + R_{max})} r \right]. \qquad (10)$$

Also, a limiting condition should be imposed on the variable $\gamma$. The variable scales mortality member in the basic equation (5) and therefore cannot exceed unity so that mortality does not exceed total number of trees, i.e.

$$\gamma \leq 1. \qquad (11)$$

Provided the normalization conditions, the resulting form of function *G(r)* will look like in fig. 2.

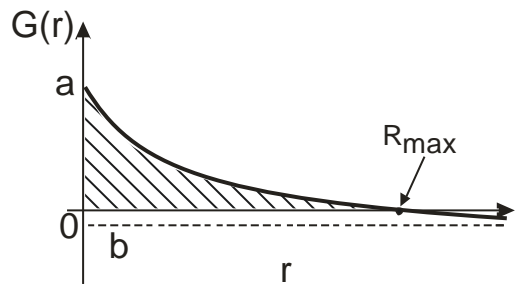

Figure 2. Form of function G(r) subject to normalization conditions (7) and (8). The hatched area is equal to unity.

Equation (10) was used for fitting real self-thinning data to test how good the model fits data and if the fitting gives realistic values of estimated parameters.

*Testing of model*

In order to perform a correct testing of the model adequate data are necessary. Figure 3 shows the course of self-thinning in the Douglas-fir stands in terms of density against mean diameter at breast height. It is obvious that earlier stages of self-thinning have some flatter decrease of density than later stages. The less the initial density the longer is the flatter stage. Especially, this can be seen on the example of Clemons and Skykomish experiments where mean diameter increased by one half with no significant mortality in the stand.

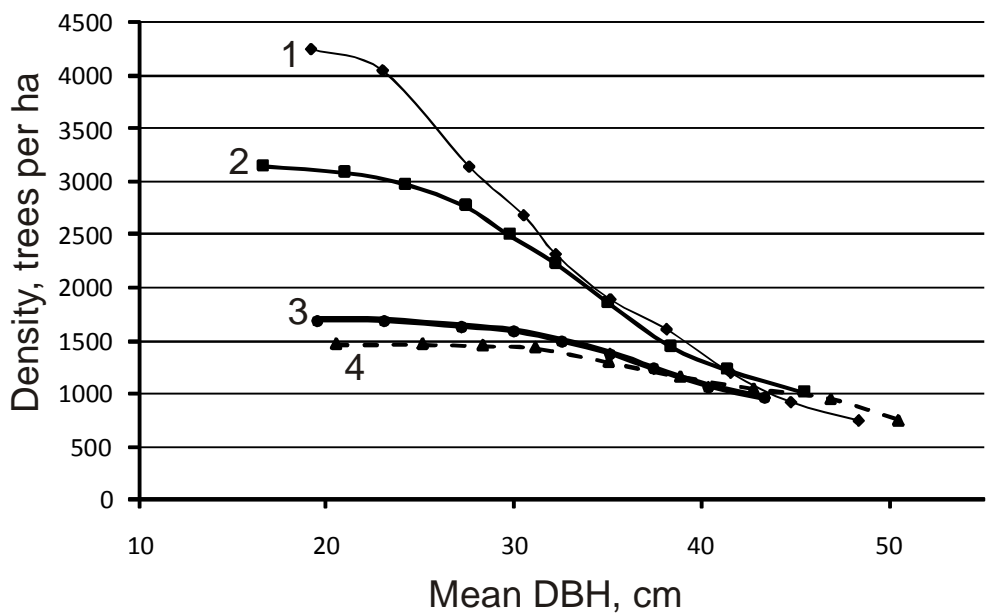

Figure 3. The course of self-thinning in Douglas-fir stands: 1 – Hoskins (Marshall, Curtis, 2001), 2 – Iron Creek (Curtis, Marshall, 2009), 3 – Clemons and 4 – Skykomish (King et al. 2002). The curves are drawn from tables given in the correspondent publications.

It is therefore reasonable to suppose that random mortality plays more important role in the earlier stages while discriminative mortality is responsible for faster drop of density at the later parts of the curves. Because the model is supposed to describe the discriminative mortality which is competition-induced it would be more adequate to use those parts of curves that are thought to reflect this kind of mortality. Thus, for Hoskins experiment the first point was discarded and for other experiments first three points were discarded (see fig. 3, hollow symbols).

Another question related to the data is of the measure of horizontal size which should be used to test the model. As it has been mentioned above, trees cannot collide by stems directly; they do it through crowns or root systems. However, mass data on crown sizes, especially in dynamics, are hardly available while diameters are measured frequently and accurately. Also, widths of crowns and stem diameters are often positively correlated. Thus diameter growth may reflect competition for space though without collisions literally.

The results of model fitting of the data are presented in fig. 4 and table.

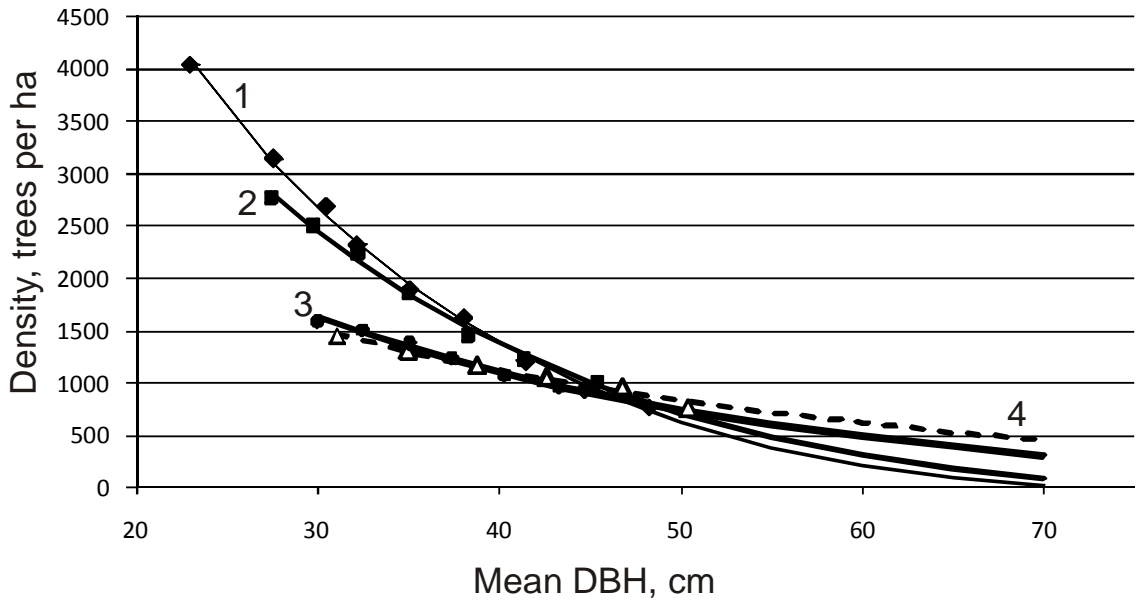

Figure 4. Results of fitting of normalized equation (5) against Douglas-fir data. Symbols are data, numbers indicate lines of regression: Hoskins (♦, 1), Iron Creek (■, 2), Clemons (●, 3), Skykomish (Δ, 4).

In the course of fitting, it was found that estimated variable γ tends to slightly go above unity. In accordance with the condition (11) it was deliberately set up to unity. The condition γ = 1 means that all the collided trees are dead. As a result, the equation (10) turned out to contain only two variables, $D_0$ and $R_{max}$ estimated values of which are given in the table.

Overall, in spite of substantial simplicity, normalized equation (10) successfully captures the nonlinear course of those parts of self-thinning curves in which discriminative mortality plays important role. Formally, the equation fits the available data fairly well, which is reflected in the values $R^2$ (table).

Table. Regression estimated values of variables $D_0$ and $R_{max}$ of normalized equation (5) for Douglas-fir experiments.

| Experiment | $D_0^*$, trees ha$^{-1}$ | $R_{max}^*$, cm | $R^2$ |
|---|---:|---:|---:|
| Hoskins | 28963,8 | 75,6 | 0,997 |
| Iron Creek | 22943,8 | 83,2 | 0,994 |
| Clemons | 10142,7 | 118,5 | 0,984 |
| Skykomish | 7547,9 | 154,3 | 0,976 |

* all the estimated values are significant at p < 0,05.

The values of variable $D_0$ (initial density of trees in the model) are naturally different from those initial densities that may be observed in the Douglas-fir stands. Remember that the model considers discriminative mortality that is by definition competition-induced. Discriminative stage in self-thinning usually starts not from the establishment moment giving way to random mortality. In this sense, the obtained values of $D_0$ may be assumed to be a stem density that would ensure competition and discriminative mortality from the very beginning of growth. It may be admitted that model gives realistic estimated values of such initial densities $D_0$.

While initial densities $D_0$ give a retrospective extrapolation for the self-thinning curves values of $R_{max}$ provide a prospective extrapolation. Because the fitting procedure and limitation (11) suggest that $\gamma = 1$, $R_{max}$ appears to be a sort of virtual 'goal'–mean stem diameter at which density of trees tends to zero. This tendency may be slow but nevertheless finite. The projection of the regression curves beyond the data suggests a prediction of density decrease for the finite span of diameters. According to regression estimations, values of $R_{max}$ lie within reasonable limits. Douglas-fir belongs to large tree species, the maximal recorded diameter at breast height for the species is over 4 m (Parminter, 1996), so that an estimated value of 1,5 m does not look extraordinary.

Reviewing the model suggested and its analysis one can come to a thought that the model development is very complicated for no real reason. In fact, it is known from long ago that–from the viewpoint of data fitting–a polynomial will work very well for most data. For example, the considered self-thinning curves may be fitted by a simple polynomial of second order with much better fitting quality. However, it is also known that the fundamental problem of such a fitting is that the received parameters are tractable neither biologically no physically in most cases. Thus the model considered here suggests a mathematical form in which every input parameter has a tractable meaning.

Another goal of the model development was introduction of spatial dynamics into consideration of self-thinning. In analytical models of self-thinning, it is quite seldom that changes of tree spatial arrangements are taken into account. Mortality of trees in dense stands is largely space-dependent. Mortality depends on space arrangement and mortality changes space arrangement of trees. It is in these considerations of spatial arrangement role in self-thinning that the model was expected to contribute.

**Conclusion**

Predicting course of self-thinning is quite an important goal of research in forest science. A number of ways may be suggested to achieve it. Among them are approaches that are based on very sophisticated mathematical frameworks modeling many processes that govern local interactions and competition among trees, their neighbor-dependent growth and mortality. Mostly, such models fit well data available. A reverse side of complicated model is that they often have to incorporate hard tractable parameters.

The approach presented in this study is based on transparent balance idea that the number of trees now is what was initially minus mortality. It is then the mortality that should be modeled sufficiently right and clearly. It is widely understood that mortality in dense forest stands is largely neighbor-dependent, i.e. discriminative. Importantly, mortality is dependent on spatial arrangement but spatial arrangement is also dependent on discriminative mortality. Neighbor-dependent mortality changes spatial arrangement and these changes feed back and alter mortality itself.

It has been shown in this study that mortality may be rather simply modeled by a function that simulates the well-known pair correlation function of spatial statistics theory. The function suggests that initial arrangement of trees is close to aggregate so that mortality is high at the beginning of growth but then levels-off.

A couple of natural normalizations help to substitute less observable variables (parameters of spatial arrangement function) through a well observable variable (maximal size of trees). As a result, the final form of the model appeared to depend only on two variable, with each of which having a clear physical sense. Though rather simple, the model has been shown to well capture non-linear self-thinning dynamics of Douglas-fir stands undergoing a natural evolution of growth and discriminative mortality.

**Acknowledgements**

The authors acknowledge support from Siberian Federal University. Also, we are grateful to Dr. O.P. Secretenko who read the manuscript and provided useful criticism on the research. The study was in part supported by the Russian Foundation for Basic Research, research grant 16-45-240411 'Influence of regional climate change on substance turnover in natural protected landscapes of Middle Siberia: a diagnosis based on use of natural (Be-7) and anthropogenic (Cs-137) radionuclide markers'.